\documentclass{article}
\usepackage{amsmath, amssymb, amsfonts}
\title{ A Vaidya-type spacetime with no singularities } 
\author{Hristu Culetu, \\Ovidius University, Department of Physics and Electronics, \\ Bld. Mamaia  124, 900527 Constanta, Romania\footnote{electronic address: hculetu@yahoo.com}}

\begin{document}
\numberwithin{equation}{section}
\pagenumbering{arabic}
\maketitle
\newcommand{\fv}{\boldsymbol{f}}
\newcommand{\tv}{\boldsymbol{t}}
\newcommand{\gv}{\boldsymbol{g}}
\newcommand{\OV}{\boldsymbol{O}}
\newcommand{\wv}{\boldsymbol{w}}
\newcommand{\WV}{\boldsymbol{W}}
\newcommand{\NV}{\boldsymbol{N}}
\newcommand{\hv}{\boldsymbol{h}}
\newcommand{\yv}{\boldsymbol{y}}
\newcommand{\RE}{\textrm{Re}}
\newcommand{\IM}{\textrm{Im}}
\newcommand{\rot}{\textrm{rot}}
\newcommand{\dv}{\boldsymbol{d}}
\newcommand{\grad}{\textrm{grad}}
\newcommand{\Tr}{\textrm{Tr}}
\newcommand{\ua}{\uparrow}
\newcommand{\da}{\downarrow}
\newcommand{\ct}{\textrm{const}}
\newcommand{\xv}{\boldsymbol{x}}
\newcommand{\mv}{\boldsymbol{m}}
\newcommand{\rv}{\boldsymbol{r}}
\newcommand{\kv}{\boldsymbol{k}}
\newcommand{\VE}{\boldsymbol{V}}
\newcommand{\sv}{\boldsymbol{s}}
\newcommand{\RV}{\boldsymbol{R}}
\newcommand{\pv}{\boldsymbol{p}}
\newcommand{\PV}{\boldsymbol{P}}
\newcommand{\EV}{\boldsymbol{E}}
\newcommand{\DV}{\boldsymbol{D}}
\newcommand{\BV}{\boldsymbol{B}}
\newcommand{\HV}{\boldsymbol{H}}
\newcommand{\MV}{\boldsymbol{M}}
\newcommand{\be}{\begin{equation}}
\newcommand{\ee}{\end{equation}}
\newcommand{\ba}{\begin{eqnarray}}
\newcommand{\ea}{\end{eqnarray}}
\newcommand{\bq}{\begin{eqnarray*}}
\newcommand{\eq}{\end{eqnarray*}}
\newcommand{\pa}{\partial}
\newcommand{\f}{\frac}
\newcommand{\FV}{\boldsymbol{F}}
\newcommand{\ve}{\boldsymbol{v}}
\newcommand{\AV}{\boldsymbol{A}}
\newcommand{\jv}{\boldsymbol{j}}
\newcommand{\LV}{\boldsymbol{L}}
\newcommand{\SV}{\boldsymbol{S}}
\newcommand{\av}{\boldsymbol{a}}
\newcommand{\qv}{\boldsymbol{q}}
\newcommand{\QV}{\boldsymbol{Q}}
\newcommand{\ev}{\boldsymbol{e}}
\newcommand{\uv}{\boldsymbol{u}}
\newcommand{\KV}{\boldsymbol{K}}
\newcommand{\ro}{\boldsymbol{\rho}}
\newcommand{\si}{\boldsymbol{\sigma}}
\newcommand{\thv}{\boldsymbol{\theta}}
\newcommand{\bv}{\boldsymbol{b}}
\newcommand{\JV}{\boldsymbol{J}}
\newcommand{\nv}{\boldsymbol{n}}
\newcommand{\lv}{\boldsymbol{l}}
\newcommand{\om}{\boldsymbol{\omega}}
\newcommand{\Om}{\boldsymbol{\Omega}}
\newcommand{\Piv}{\boldsymbol{\Pi}}
\newcommand{\UV}{\boldsymbol{U}}
\newcommand{\iv}{\boldsymbol{i}}
\newcommand{\nuv}{\boldsymbol{\nu}}
\newcommand{\muv}{\boldsymbol{\mu}}
\newcommand{\lm}{\boldsymbol{\lambda}}
\newcommand{\Lm}{\boldsymbol{\Lambda}}
\newcommand{\opsi}{\overline{\psi}}
\renewcommand{\tan}{\textrm{tg}}
\renewcommand{\cot}{\textrm{ctg}}
\renewcommand{\sinh}{\textrm{sh}}
\renewcommand{\cosh}{\textrm{ch}}
\renewcommand{\tanh}{\textrm{th}}
\renewcommand{\coth}{\textrm{cth}}

\begin{abstract}
A regular Vaidya-type line-element is proposed in this work. The mass function depends both on the temporal and the spatial coordinates. The curvature invariants and the source stress tensor $T^{a}_{~b}$ are finite in the whole space. The energy conditions for  $T^{a}_{~b}$ are satisfied if $k^{2}<2vr$, where $k$ is a positive constant and $v,r$ are  coordinates. It is found that the radial pressure has a maximum very close to $r = 2m~ (r>2m), v = 2m$. The energy crossing a sphere of constant radius is akin to Lundgren-Schmekel-York quasilocal energy. The Newtonian acceleration of the timelike geodesics has an extra term (compared to the result of Piesnack and Kassner) which leads to rejecting effects. 
 \end{abstract}
 
 \section{Introduction}
  As a nonstatic generalization of the Schwarzschild geometry, the Vaidya solution \cite{PV} represents an imploding (exploding) null dust fluid with spherical symmetry and describes the state of a collapsing star (the exterior metric \cite{NM, WW, GG}). Whereas Wang and Wu generalize the Vaidya geometry including a mass function depending not only on the temporal variable but also on the radial one, Govinder and Govender \cite{GG} studied a null radiation-string fluid model where the end state is a naked singularity (see also \cite{DJ}). The presence of the string matter contributes to the appearance of the naked singularity.   
		
	Booth and Martin \cite{BM} proposed a geometrical measure of the distance between the apparent and the event horizon (EH) for a dynamical black hole (BH) and investigated them for the Vaidya spacetime. They located the two horizons and their separation has been determined using the timelike geodesics normal to the apparent horizon. 
	
	Mkenyeleye et al. \cite{MGM} showed that there exists classes of the generalized Vaidya mass function (or the Misner-Sharp mass) for which the collapse terminates with a locally naked central singularity. For the special sub-classes of self-similar Vaidya geometry, their mass function $m(v,r)$ depends only on the advanced null coordinate $v$, with $m(v) = 0$ for $v<0$, $m(v) \propto v$ for $0\leq v \leq T$, and $m(v) = M_{0}$ for $v>T$, where $M_{0}$ is the Schwarzschild constant mass. Their mass function is a non-negative increasing function of $v$ for imploding radiation. In the authors' view, the generalized Vaidya metric is more realistic than dust-like matter or perfect fluid during the later stages of the collapse of a massive star.
	
	Piesnack and Kassner \cite{PK} used the ingoing Vaidya geometry for a nonrotating uncharged BH emitting Hawking radiation. They stated that a particle released from rest closed enough to the apparent horizon is repelled and may escape to infinity, an effect which is missing in Schwarzschild geometry. However, the presence of negative energy density alone does not explain the repulsion, although it may be a necessary condition for that.
	
	Motivated by the previous studies on a generalized Vaidya geometry, with a temporal and spatial dependence of the mass function, we propose a similar metric but with no singularities and where the Ricci and Kretschmann scalars are regular in the whole space. Moreover, the energy conditions for the source energy-momentum tensor are generally obeyed. We found the expression of the energy crossing a surface of constant radius resembles the Lundgren-Schmekel-York (LSY) quasilocal energy \cite{LSY}. However, after a power series expansion of the energy flow $W$, the 2nd term has a minus sign, in contrast with the authors of \cite{LSY}. That is due to a different power in the expression of $W(r)$.
		
	The paper is organized as follows. In Sec.2 we present the spacetime in terms of a general mass function, showing the components of the Einstein tensor in terms of the mass function $M(v,r)$. Sec.3 is devoted to the regular Vaidya geometry to which the curvature invariants and kinematical quantities associated to a typical observer are computed. The line-element becomes regular thanks to an exponential factor depending both on $v$ and $r$. With that particular choice of the mass function, the source stress tensor and its energy conditions are examined in Sec.4. The simple mathematical form of the regular energy density and pressures of the imperfect fluid is also emphasized. 
	
	The gravitational energy crossing a sphere of constant radius is calculated in Sec.5. It is independent on the constant parameter $k$ from the expression of the mass function. The radial timelike geodesics are investigated in Sec.6 and a comparison with the results from Ref.\cite{PK} is shown. Sec.7 is dealing with the microscopic situation, where our main metric is to be applied, a case where the constant mass $m$ represents an elementary particle and the temporal variable signifies the duration of the performed measurement. For the constant $k$ from the metric we have chosen here the Compton wavelength associated to the particle of mass $m$.
	Finally, some concluding remarks are given in Sec.8.
	
	\section{Generalized Vaidya geometry}
	We start with a spherically-symmetric spacetime in advanced time-coordinate $v$, having the form \cite{GG, GK, BKS}
		 \begin{equation}
  ds^{2} = -\left[1 - \frac{2M(v,r)}{r}\right]dv^{2} + 2dvdr + r^{2} d \Omega^{2}, 
 \label{2.1}
 \end{equation}
 where $M(v,r)$ is the mass function, $v$ is the Vaidya ingoing null coordinate and $d \Omega^{2}$ stands for the metric of the unit two-sphere. 

 The Einstein equations 
 		 \begin{equation}
		G_{ab} \equiv R_{ab} - \frac{1}{2}g_{ab}R = 8\pi T_{ab}
 \label{2.2}
 \end{equation}
give us the connection between the geometry (the Einstein tensor) and matter (the stress tensor $T_{ab}$). For the metric (2.1) we calculate the following non-zero mixed components of $G_{ab}$ with the help of Maple package
  \begin{equation}
	G^{v}_{~v} = G^{r}_{~r} = -\frac{2}{r^{2}} M' ,~~~G^{r}_{~v} = \frac{2}{r^{2}} \dot{M} ,~~~G^{\theta}_{~\theta} = G^{\phi}_{~\phi} = -\frac{1}{r} M'' 
 \label{2.3}
 \end{equation}
where $M' \equiv \partial M(v,r)/\partial r$ and $\dot{M} \equiv \partial M(v,r)/\partial v$. As the source of the geometry (2.1) we employ an energy-momentum tensor corresponding to an imperfect fluid with energy flux \cite{GRM, HC1, HC2}
	  \begin{equation}
	 T_{ab} = (p_{t} + \rho) u_{a} u_{b} + p_{t} g_{ab} + (p_{r} - p_{t}) n_{a}n_{b} +  u_{a} q_{b} + u_{b} q_{a},
 \label{2.4}
 \end{equation}
where $\rho$ is the energy density of the fluid, $p_{r}$ is the radial pressure, $p_{t}$ are the transversal (tangential) pressures, $q^{a}$ is the energy flux 4-vector and $n^{a}$ is a unit spacelike vector orthogonal to $u^{a}$. We have $u_{a} n^{a} = 0,~u_{a} u^{a} = -1,~n_{a} n^{a} = 1$ and $u_{a} q^{a} = 0$. In the geometry (2.1) one chooses a 4-velocity of the form
	  \begin{equation}
		u^{a} = \left(1, -\frac{M}{r}, 0, 0\right), 
 \label{2.5}
 \end{equation}
whence, keeping in mind the above relations
	  \begin{equation}
		n^{a} = \left(1, 1 -\frac{M}{r}, 0, 0\right).
 \label{2.6}
 \end{equation}
One observes that we obtained from (2.1) and (2.2) four equations and five unknown functions, namely $\rho, p_{r}, p_{t}, q$ and $M$. Usually, an equation of state connecting the energy density and pressure is introduced for to solve the system of equations. 

 We propose to employ a different route for to achieve that goal: one looks for an appropriate expression of the mass function in order to obtain a regular Vaidya-type geometry. 
	
	\section{Regular Vaidya spacetime}
	One notices from Eqs.(2.3) that, once we know the expression of the mass function $M(v,r)$ all the components of the stress tensor can be determined in terms of $v$ and $r$ and  constants of integration. To make the spacetime regular we take advantage of a previus paper \cite{HC4} (see also \cite{HC5}) where we introduced an exponential factor $e^{-\kappa/r}$, $\kappa$ being a positive constant, for to render the Schwarzschild metric regular. Because the Vaidya geometry is nonstatic, we have to adjust the previous exponential factor in order to contain the variable $v$. 
	
	We propose the following spacetime to be valid in the exterior of a collapsing star
	 \begin{equation}
  ds^{2} = -\left(1 - \frac{2m}{r} e^{-\frac{k^{2}}{vr}}\right)dv^{2} + 2dvdr + r^{2} d \Omega^{2}, 
 \label{3.1}
 \end{equation}
where $m$ is a constant mass, $e = 2.718...$ is Euler's number, with $log~e = 1$, where $log$ stands for the natural logarithm and $k$ represents a positive constant length. It is clear from the above equation that 
	 \begin{equation}
	M \equiv M(v,r) = m e^{-\frac{k^{2}}{vr}}
 \label{3.2}
 \end{equation}
plays the role of the mass function. The special choice $k = 0$ leads to the Eddington-Finkelstein form of the Schwarzschild line-element. Moreover, $M$ is an increasing function, both of $v$ and of $r$, and varies from zero to $m$.
The metric coefficient $f(v,r) = -g_{vv}= 1-(2m/r) e^{-k^{2}/vr}$ is regular for any $v,r$ and (3.1) becomes Schwarzschild line-element with the mass $m$ for $v \rightarrow \infty$. Following the authors of \cite{MGM, JDJ, BR, JW}, we assume the mass function $M(v,r)$ to be zero for $v<0$, when the spacetime is Minkowskian, but for $v\geq 0$ it is given by the expression (3.2). As Jhingan et al. \cite{JDJ} have observed, for a realistic evolution of the physical system it is necessary that there is no energy flux  across v = 0 and that the mass function has the same value at it, when approached from either side. Both conditions are obeyed for our mass function (3.2) and so our choice when $v<0$ is justified. It is worth noticing that $M$ has the same dependence as a function of $v$ or $r$. It is vanishing for $v\rightarrow 0$ or $r\rightarrow 0$, and so there is a continuity with the case $v<0$, where $M$ is considered to be zero. 

It is clear that when $v>0$ and $r>2m$, the function $f(v,r)$ is positive. Therefore, from now on, we study only this region. Noting that we have no signature flip of $f(v,r)$ there. We let the region $v>0,~0<r<2m$ (when $v$ may become spacelike and $r$-timelike) for a future work.

 Let us observe that we have a natural length at our disposal to insert it instead of $k$ in the metric (3.1): the Schwarzschild radius of the constant mass $m$. This particular situation will be especially useful when we will examine the extrema of the energy density and pressures associated to the imperfect fluid.

The scalar curvature corresponding to the spacetime (3.1) is given by
	 \begin{equation}
	R^{a}_{~a} = \frac{2mk^{4}}{v^{2}r^{5}}e^{-\frac{k^{2}}{vr}},
 \label{3.3}
 \end{equation}
One sees that it tends to zero both for $v \rightarrow 0$ and for $v \rightarrow \infty,~r\rightarrow \infty$, because of the exponential factor. For the Kretschmann scalar $K = R^{abcd}R_{abcd}$ one obtains
	 \begin{equation}
	K = \frac{48m^{2}}{r^{6}}\left(1-\frac{2k^{2}}{vr} +\frac{2k^{4}}{v^{2}r^{2}} -\frac{2k^{6}}{3v^{3}r^{3}} +\frac{k^{8}}{12v^{4}r^{4}}\right)e^{-\frac{2k^{2}}{vr}}
 \label{3.4}
 \end{equation}
where the indices $a,b,c,d$ label ($v,r,\theta,\phi$). Noting that, with $r$-constant and $v\rightarrow \infty$, $K$ acquires the Schwarzschild value $48m^{2}/r^{6}$. We have again a finite $K$ for any values of $v$ and $r$, thanks to the exponential factor. 
 For to investigate the kinematical quantities, we need the velocity vector field (2.5) which becomes now
   \begin{equation}
	 u^{a} = \left(1, - \frac{m}{r}e^{-\frac{k^{2}}{vr}}, 0, 0 \right),	~~~~~~u^{a}u_{a}= -1.
 \label{3.5}
 \end{equation} 
	The vector $ u^{a}$ is not tangent to the geodesics, as can be seen from the nonzero components of the covarint acceleration $a^{b}= u^{a}\nabla_{a}u^{b}$ of the congruence
	     \begin{equation}
	a^{v} = \frac{m}{r^{2}} \left(1 - \frac{k^{2}}{vr}\right)e^{-\frac{k^{2}}{vr}},~~~ a^{r} = \frac{m}{r^{2}} \left(1 - \frac{k^{2}}{vr}\right)\left(1- \frac{m}{r}e^{-\frac{k^{2}}{vr}}\right)e^{-\frac{k^{2}}{vr}}.
 \label{3.6}
 \end{equation}
The proper acceleration reads
  \begin{equation}
\sqrt{a^{b}a_{b}}	= \frac{m}{r^{2}} |1 - \frac{k^{2}}{vr}|e^{-\frac{k^{2}}{vr}}
 \label{3.7}
 \end{equation}
which equals $a^{v}$ in the region $vr>k^{2}$. The scalar expansion of the congruence is given by
  \begin{equation}
\Theta \equiv \nabla_{a}u^{a} = -\frac{m}{r^{2}} \left(1 + \frac{k^{2}}{vr}\right)e^{-\frac{k^{2}}{vr}},
 \label{3.8}
 \end{equation}
	that is negative for any values of the variables, as expected for ingoing matter.
	
	\section{Imperfect fluid stress tensor}
	Having known the expression of the mass function (3.2) we find the Einstein tensor from the Eqs.(2.3). Then, from the Einstein equations $G^{a}_{~b} = 8\pi T^{a}_{~b}$ and the form (2.4) of the stress tensor, one obtains:\\
	- from the $vv$-component
	  \begin{equation}
  \frac{M}{r}(\rho + p_{r}) - \rho + q_{v} - \frac{M}{r} - \left(1 - \frac{M}{r}\right)q^{v} = -\frac{k^{2}M}{4\pi vr^{4}}. 
 \label{4.1}
 \end{equation}
- from the $rr$-component
	  \begin{equation}
  -\frac{M}{r}(\rho + p_{r}) + p_{r} - \frac{M}{r}q_{r} + q^{r}  = -\frac{k^{2}M}{4\pi vr^{4}}. 
 \label{4.2}
 \end{equation}
- from the $rv$-component
	  \begin{equation}
  \frac{M}{r}\left(1 - \frac{M}{r}\right)(\rho + p_{r}) - \frac{M}{r}q_{v} - \left(1 - \frac{M}{r}\right) q^{r}  = \frac{k^{2}M}{4\pi v^{2}r^{3}}. 
 \label{4.3}
 \end{equation}
- from the $vr$-component
	\begin{equation}
	\rho + p_{r} + q_{r} + q^{v} = 0.
 \label{4.4}
 \end{equation}
- from the $\theta \theta$-component
		\begin{equation}
		p_{t} = \frac{k^{2}M}{4\pi vr^{4}}\left(1 - \frac{k^{2}}{2vr}\right).
 \label{4.5}
 \end{equation}
From $u_{a} q^{a} = 0$ we get, in terms of $q^{v}$
		\begin{equation}
		q_{v} = \frac{M}{r}q^{v},~~~q^{r} = \left(1 - \frac{M}{r}\right)q^{v},~~~q_{r} = q^{v}.
 \label{4.6}
 \end{equation}
Therefore, the above equations yield
		\begin{equation}
		\begin{split}
	  \frac{M}{r}(\rho + p_{r}) - \rho - \left(1 - \frac{2M}{r}\right)q^{v} = -\frac{k^{2}M}{4\pi vr^{4}}\\ 
	  -\frac{M}{r}(\rho + p_{r}) + p_{r} + \left(1 - \frac{2M}{r}\right)q^{v} = -\frac{k^{2}M}{4\pi vr^{4}}\\ 
     \frac{M}{r}\left(1 - \frac{M}{r}\right)(\rho + p_{r}) - \left(1 - \frac{2M}{r} + \frac{2M^{2}}{r^{2}}\right) q^{v}  = \frac{k^{2}M}{4\pi v^{2}r^{3}}\\ 
   \rho + p_{r} + 2q^{v} = 0.
 \label{4.7}
 \end{split}
 \end{equation}
From the last Eq.(4.7) we have $q^{v} = -(\rho + p_{r})/2$. The first and second equations (4.7) give us the same relation
		\begin{equation}
	\rho - p_{r} = \frac{k^{2}M}{2\pi vr^{4}},
 \label{4.8}
 \end{equation}
but that one concerning the component ($rv$) leads to 
	\begin{equation}
	\rho + p_{r} = \frac{k^{2}M}{2\pi v^{2}r^{3}}.
 \label{4.9}
 \end{equation}
One sees that $\rho$ and $p_{r}$ are easily obtained from (4.8) and (4.9), with $M$ known from (3.2)
	\begin{equation}
 \rho = \frac{mk^{2}}{4\pi v^{2}r^{3}}\left(1 + \frac{v}{r}\right)e^{-\frac{k^{2}}{vr}}, ~~~p_{r} = \frac{mk^{2}}{4\pi v^{2}r^{3}}\left(1 - \frac{v}{r}\right)e^{-\frac{k^{2}}{vr}}.
 \label{4.10}
 \end{equation}
However, we have no an extra equation where to insert the expressions (4.10) and so to fix the constant $k$ (eventually in terms of $m$), for any $v$ and $r$ from their domain of variation. Hence, the solution of Einstein equations is valid for any length $k$. That conclusion is generated by the fact that the first and second equations (4.7) gives us the same relation (4.8). 
	
The nonzero components of the energy flux are given by
	\begin{equation}
	 q^{v} = -\frac{mk^{2}}{4\pi v^{2}r^{3}}e^{-\frac{k^{2}}{vr}},~~~ q^{r} = \left(1 - \frac{m}{r}e^{-\frac{k^{2}}{vr}}\right)q^{v},~~~with ~~ q \equiv \sqrt{q^{a}q_{a}} = -q^{v}. 
 \label{4.11}
 \end{equation}
In other words, $q = T^{r}_{~v}$, namely the energy flux equals the only out of diagonal component of the stress tensor. 

One observes that, with an arbitrary $k$, $\rho, p_{r}, p_{t}, q$ do not depend on the Newton constant $G$ (even though we started with the gravitational equations), because we have one $G$ in $8\pi G/c^{4}$ and the other in $2Gm/c^{2}$. The situation changes only for the special value $k = 2m$ (or a function of $m$).

 As long as the energy conditions for $T^{a}_{~b}$ are concerned, one notices from (4.10) and the expression of $p_{t}$ from (4.5) that the weak energy condition (WEC) ($\rho>0, \rho +p_{r}>0, \rho +p_{t}>0$), the dominant energy condition (DEC) ($\rho>|p_{r}|, \rho>|p_{t}|$), the null energy condition (NEC) ( $\rho +p_{r}>0, \rho +p_{t}>0$) and the strong energy condition (SEC) ($\rho +p_{r}>0, \rho +p_{t}>0, \rho +p_{r} + 2p_{t}>0$) are satisfied, if $k^{2}/2vr <1$ (when transversal pressures are positive); otherwise, the result depends on the value of $k$.

Let us find now the extrema of the quantities $\rho, p_{r}, p_{t}, q$. To find them, we need the critical points. Those are obtained from the roots of the first partial derivatives w.r.t. $v$ and $r$. 

 For the energy density, one obtains from (4.10) 
	\begin{equation}
	\begin{split}
	\frac{\partial \rho}{\partial v} = 0,~~~~~\rightarrow ~~~rv^{2} - k^{2}v + 2r^{2}v - k^{2}r = 0\\  \frac{\partial \rho}{\partial r} = 0,~~~~~\rightarrow ~~~4rv^{2} - k^{2}v + 3r^{2}v - k^{2}r = 0.
 \label{4.12}
\end{split}
 \end{equation}
The two equations from (4.12) yield $vr(r + 3v) = 0$, with no solutions.

For the radial pressure we get
 	\begin{equation}
	\begin{split}
	\frac{\partial p_{r}}{\partial v} = 0,~~~~~\rightarrow ~~~rv^{2} - k^{2}v - 2r^{2}v + k^{2}r = 0\\  \frac{\partial p_{r}}{\partial r} = 0,~~~~~\rightarrow ~~~4rv^{2} - k^{2}v - 3r^{2}v + k^{2}r = 0.
 \label{4.13}
\end{split}
 \end{equation}
The two equations give us the solution $v = k\sqrt{2/15},~r = k\sqrt{6/5}$, that represents the critical point. We mention that for the special value $k = 2m$ one obtains $r>2m$, which assures that $f(v,r)>0$. To see whether the critical point is an extremum, we need the values of the 2nd derivatives of $p_{r}$ w.r.t. $v$ and $r$ at the critical point. The calculation yields
  	\begin{equation}
	\frac{\partial^{2} p_{r}}{\partial v^{2}} \approx -33.09 \frac{m}{k^{5}},~~~ \frac{\partial^{2} p_{r}}{\partial r^{2}} \approx -147.34 \frac{m}{k^{5}},~~~\frac{\partial^{2} p_{r}}{\partial r \partial v} \approx -19.24 \frac{m}{k^{5}}.
 \label{4.14}
 \end{equation}
 By means of the Hessian matrix we infer that the above critical point is a local maximum. The maximum value of the radial pressure will be $p_{r,max} = \frac{25}{24\pi} \sqrt{\frac{5}{6}}~e^{-\frac{5}{2}} \frac{m}{k^{3}} \approx 2.78\cdot 10^{-2} m/k^{3}$.

  One observes that the expressions (4.5), (4.10) and (4.11), originating from Einstein's equations with the choice (3.2) of the mass function are valid for any positive constant $k$. We suggest to choose for $k$ an appropriate value in order to get plausible results. Paranjape and Padmanabhan \cite{PP} (see also \cite{RMPS, HC6}) showed that the components $T^{v}_{~v}, T^{r}_{~r}$ of the stress tensor for the Vaidya spacetime with the mass function $M(v,r)$ are of the order of $1/M^{2}$ near the apparent horizon $r = 2M(v,r(v))$. Our $p_{r,max}$ is obtained for $v = k\sqrt{2/15},~r = k\sqrt{6/5}$. That means we get $v,r$ values close to the apparent horizon if we take $k \approx 2m$. By employing this value we indeed obtain $p_{r,max} \propto 1/m^{2}$ (it is worth noting that in our expression for $p_{r,max}$ from above the exponential factor is $e^{-5/2} \approx 1/12$, so that $M$ and $m$ are of the same order of magnitude. From the arguments above it is clear that the choice $k = 2m$ seems justified.

 The value $k = 2m$ gives $p_{r,max} \approx 3.5\cdot 10^{-3}/m^{2}$. For a BH with the solar mass, one obtains $p_{r,max} \approx 10^{38} ergs/cm^{3}$, a value valid very close to $r = 2m$, i.e., at $r = 2m\sqrt{6/5}$. \\
Take now a collapsing star that leads to the formation of our Sun. Take $k = 2m$, the Sun mass $m_{s} = 3\cdot 10^{33}g$, its radius $R = 7\cdot 10^{10}cm $ and the gravitational radius $R_{s} = 3\cdot 10^{5}cm$. After, say, $v = 3\cdot 10^{16}s$ (one billion years), an observer on the surface measures an energy density $\rho\cdot \approx 0.5 \cdot 10^{-6} ergs/cm^{3}$, a tiny value compared to the previous one, close to $r = 2m$. Anyway, the observer is located far from the hypotetical horizon.

 For the transversal pressures we have
	\begin{equation}
	\begin{split}
	\frac{\partial p_{t}}{\partial v} = 0,~~~~~\rightarrow ~~~\frac{k^{2}}{vr} = 2\pm{\sqrt{2}} \\  \frac{\partial p_{t}}{\partial r} = 0,~~~~~\rightarrow ~~\frac{k^{2}}{vr} = \frac{1}{2}(7 \pm \sqrt{17}).
 \label{4.15}
\end{split}
 \end{equation}
One observes that the two equations (4.15) are not compatible, whence one concludes that $p_{t}$ has no critical points. 

It can be shown that the energy flux $q$ has no critical points, too.

\section{Total energy flow}
Having now the components of the stress tensor and the basic physical quantities associated to it, our next task is to compute the total energy crossing a sphere of radius $r = r_{0}$=const. \cite{HC1}, with $r_{0}>2m$.
 \begin{equation}
 W = \int{T^{a}_{~b}u^{b}n_{a}\sqrt{-\gamma}}dv~d\theta~d\phi,
 \label{5.1}
 \end{equation}
where $u^{b}$ and $n^{b}$ are given by (2.5) and (2.6), respectively, with $M$ from (3.2). To find $\gamma$ we remove the $r$-line and $r$- column in $g_{ab}$ and obtain $\gamma = -(1 - \frac{2m}{r}e^{-\frac{k^{2}}{vr}})r^{4} sin^{2}\theta$. With $T^{r}_{~v}$ from (4.3), Eq. (5.1) yields
 \begin{equation}
 W = \frac{mk^{2}}{r_{0}}\int^{\infty}_{0} \frac{1}{v^{2}}~e^{-\frac{k^{2}}{vr_{0}}}\sqrt{1 - \frac{2m}{r_{0}}e^{- \frac{k^{2}}{vr_{0}}}}dv    
 \label{5.2}
 \end{equation}
We change in (5.2) the variable of integration to $h = 1 - \frac{2m}{r_{0}}e^{- \frac{k^{2}}{vr_{0}}}$, which is a function of $v$. We get
 \begin{equation}
W = -\frac{r_{0}}{2}\int^{1 - 2m/r_{0}}_{1} \sqrt{h}dh .       
 \label{5.3}
 \end{equation}
 One finally obtains  
 \begin{equation}
W = -\frac{r_{0}}{3}h^{3/2}~|_{1}^{1 - 2m/r_{0}} = \frac{r_{0}}{3} \left[1 - \left(1 - \frac{2m}{r_{0}}\right)^{3/2}\right],~~~~r_{0}\geq 2m.
 \label{5.4}
 \end{equation}
Let us observe that energy flow $W$ from (5.4) does not depend on the constant $k$. It resembles the quasilocal energy of Lundgren et al. \cite{LSY} for a Schwarzschild BH. However,   when $r_{0}>>2m$, a power series expansion in (5.4) gives us
 \begin{equation}
W \approx m - \frac{m^{2}}{2r_{0}}.
 \label{5.5}
 \end{equation}
It is worth noting the minus sign on the r.h.s. of (5.5), in contrast with the result of Lundgren et al. (their Eq.(2)). The difference comes from the power $3/2$ from (5.4), compared to the power $1/2$ from \cite{LSY}. Anyway, the second term from (5.5) is much less than the first in the approximation used (it tends to zero when $r_{0} \rightarrow \infty$; that is, (5.5) becomes $W = m$). Because $W(2m) = 2m/3$, one concludes that $2m/3 \leq W \leq m$.

\section{Timelike geodesics}
As a next step, we plan to find the equations of motion for a test particle radially falling in the spacetime (3.1). From the Lagrangean (see \cite{PK})
  \begin{equation}
L \equiv - \frac{ds^{2}}{d\tau^{2}} = (1 - \frac{2m}{r} e^{-\frac{k^{2}}{vr}})\dot{v}^{2} - 2\dot{v} \dot{r} - r^{2}(\dot{\theta}^{2} + sin^{2}\theta~ \dot{\phi}^{2}),        
 \label{6.1}
 \end{equation}
we should infer the geodesic equations, with $\theta, \phi = const.$ In this section, the dot means a derivative w.r.t. the proper time $\tau$ of the particle. The Euler-Lagrange equations read
  \begin{equation}
	\frac{\partial L}{\partial v} - \frac{d}{d\tau}\frac{\partial L}{\partial \dot{v}} = 0,~~~~\frac{\partial L}{\partial r} - \frac{d}{d\tau}\frac{\partial L}{\partial \dot{r}} = 0
 \label{6.2}
 \end{equation}
Keeping in mind that $L = 1$ and using (6.2), one concludes that
  \begin{equation}
	\ddot{v} + \frac{m}{r^{2}}\left(1 - \frac{k^{2}}{vr}\right)e^{-\frac{k^{2}}{vr}} \dot{v}^{2} = 0,~~~ 
	\ddot{r} + \frac{m}{r^{2}}\left(1 - \frac{k^{2}}{vr} + \frac{k^{2}}{v^{2}}\dot{v}^{2}\right)e^{-\frac{k^{2}}{vr}} = 0
 \label{6.3}
 \end{equation}
For a comparison with Eq.19 from \cite{PK}, we write down the 2nd equation (6.3) for the Newtonian acceleration as
  \begin{equation}
	\ddot{r} = -\frac{M(v,r)}{r^{2}} - \frac{1}{r}\dot{v}^{2}\frac{\partial M(v,r)}{\partial v} + \frac{v}{r^{2}}\frac{\partial M(v,r)}{\partial v}
 \label{6.4}
 \end{equation}
One observes that the 1st two terms from the r.h.s. of (6.4) are identical with the corresponding ones from \cite{PK}, Eq.19. However, in their article the 2nd term is positive due to the negative factor $r'_{s}(v)$. In our work, the 2nd term is negative but the 3rd is positive. Therefore, this 3rd term leads to repulsive effects. For example, if $\dot{v}^{2} - v/r <0$, the last two terms may have together a repulsive contribution, when $v/r>1$. That is in accordance with $p_{r}<0$ when $v/r>1$ (see Eq.4.2); indeed, a negative pressure leads to repulsion. 

\section{Application in microphysics}
We applied so far the geometry (3.1) for a collapsing star. Therefore, the constant distance $k$ received the special value $k = 2m$, namely the gravitational radius of mass $m$. But  BHs with a mass less than the Planck mass $m_{P} \approx 10^{-5}$ grams cannot arise from direct gravitational collapse; they might be the consequence of density fluctuations in some violent subatomic process \cite{PK} (see also \cite{CN}). But in microphysics we have a special choice for the constant length: $k = \hbar/mc \equiv \lambda$, namely, the reduced Compton wavelength associated to the mass $m$. Consequently, the spacetime (3.1) becomes 
  \begin{equation}
  ds^{2} = -\left(1 - \frac{2m}{r} e^{-\frac{1}{m^{2}vr}}\right)dv^{2} + 2dvdr + r^{2} d \Omega^{2}.	
 \label{7.1}
 \end{equation}
Of course, the geometry has not changed, but the magnitude of the parameters did, when the arbitrary $k$ is replaced by $1/m$. It is understood that $m$ is the mass of an elementary particle (the Compton wavelength is greater than its radius). In addition, we assume a different physical meaning of the variable $v$: it represents the duration of some measurement performed upon the physical system into consideration, with the sense that no measurement signifies infinite duration. 

From (7.1) one observes that $m = 0$ leads to $f = 1$, as expected (flat space). We have again $r>2m$, to get positive $f(v,r)$. Moreover, for $v \rightarrow \infty$  we obtain the static (Schwarzschild) situation. On the contrary, $v \rightarrow 0$ gives again the flat space, as for $m = 0$. That means the shorter the measurement duration, the weaker the effect of the gravitational source $m$ is. A similar idea has been also developed in \cite{HC3} for a nonstatic Schwarzschild-like geometry. It is worth noting that the expression (5.4) for the energy flow $W$ is also valid here because it is independent on the constant $k$ (the Compton wavelength in this section).

To get sensible effects, it is clear that the variables $v, r$ from (7.1) should take small values. To see this, let us write the expression of the energy density of the imperfect fluid including all fundamental constants
  \begin{equation}
	8\pi \rho = \frac{2\hbar^{2}}{mcvr^{4}}\left(1 + \frac{r}{cv}\right)e^{-\frac{\hbar^{2}}{m^{2}c^{3}vr}}.
 \label{7.2}
 \end{equation}
 The expression (7.2) depends on the fundamental constants $\hbar$ and $c$, with no dependence on $G$. It is semiclassical and $\rho$ is vanishing for $\hbar = 0$ or $m \rightarrow 0$. 

Let us consider $m = m_{e} \approx 10^{-27}g$ (the electron mass). The corresponding Compton wavelength is $\lambda_{e} \approx 3\cdot 10^{-11}$ cm. If one takes $r = 10^{-10}$ cm ($r>2m$) and $v = 10^{-22}$s, the exponential factor from the expression of $f(v,r)$ is $e^{-3}$, which diminishes the mass function $M(v,r)$. However, the electron gravitational radius is too small and $f \approx 1$. In contrast, the situation is very different as far as the energy density of the fluid is concerned. Using the same values of $v$ and $r$ as above, one obtains from (7.2) that $\rho \approx 4\cdot 10^{23} ergs/cm^{3} = 0.4 ergs/A^{3}$, which is no longer negligible. That means at a distance $10^{-10}$ cm from an electron, a measurement on the energy density  performed in $10^{-22}$ s would give the above value for $\rho$.\\ 
 For the macroscopic case (Sec.4), we got $p_{r,max} \approx 2.78\cdot 10^{-2} m/k^{3}$. We replace here $m$ with $m_{e}$ and $k$ with $\lambda_{e}$ and obtain $p_{r,max} \approx 8.34\cdot 10^{23} ergs/cm^{3} = 0.834~ ergs/A^{3}$. That is 14 orders of magnitude less than $p_{r,max}$ for a solar mass BH. 

   Let us make few comments regarding the above considerations. We analyzed the case of an electron as an example. But our model deals with an anisotropic fluid which we investigated initially at a macroscopic level. It is not obvious to imagine an electron as having anisotropic properties, excepting the situation when anisotropy is small, i.e. the pressures (and eventually the energy density of the fluid) are of the same order of magnitude (all pressures are equal for an isotropic fluid, like the perfect one). We already acquired above almost the same value for $\rho$ and $p_{r,max}$. 
	
	 Because $p_{t}$ has no extrema, it is not so easy to compare it with $p_{r}$. Anyway, $p_{r} = p_{t}$ whenever $v(r) = (2r^{2} + \lambda^{2})/4r$ (as can be seen from Eqs.4.5 and 4.10). At least on the curve $v(r)$ and around it, the fluid anisotropy is negligible. As an example, if we take $r = \lambda = 1/m$, then $v = 3/4m$. That situation gives us $p_{r} = p_{t} = m^{4}e^{-4/3}/9\pi = \hbar c~ e^{-4/3}/9\pi \lambda_{e}^{4} \approx 1.29\cdot 10^{3} ergs/cm^{3}$, a reasonable value for the considered size of $v$ and $r$.

\section{Concluding remarks}
A regular Vaidya geometry with an imperfect fluid as source of the gravitational field is investigated. The Ricci and Kretschmann scalars are regular throughout the spacetime. We proposed a mass function depending both on the radial coordinate $r$ and the null coordinate $v$. The energy density of the anisotropic fluid is positive but the transversal pressures are negative when $v<k^{2}/2r$. 

We found that all the energy conditions of the stress tensor $T^{a}_{~b}$ are fulfilled if $v>k^{2}/2r$. 
 The energy flow $W$ through a sphere of constant $r = r_{0}$ is positive and is akin to the quasilocal energy of Lundgren, Schmekel and York,Jr \cite{LSY}. However, after a power series expansion ($r_{0}>>2m$), $W(r_{0})$ has a minus sign in front of the 2nd term, in contrast with the authors of \cite{LSY}. The difference comes from the power $3/2$ from the expression of $W$. It becomes the rest energy of the mass source $m$ when $r_{0} \rightarrow \infty$. 

 We applied our regular Vaidya metric in microphysics, taking the constant $k = 1/m$, i.e. the Compton wavelength associated to the particle mass $m$ (of course, the geometry is not changed, but the magnitude of the physical parameters did). In addition, we proposed a different physical meaning of the temporal variable: it represents the duration of the measurement done upon the physical system into consideration. For instance, measurements performed in very short time interval lead to very high values of the energy density and pressures. The mathematical expressions of the energy density and pressures do not depend on the Newton constant $G$ but, of course, they will depend on the Planck constant, through the Compton wavelength. Moreover, when $\hbar = 0$ or $m \rightarrow 0$, they are vanishing. \\

\textbf{Acknowledgements}
 
I am grateful to the one of the anonymous referees for useful suggestions and comments which considerably improved the quality of the manuscript.

\end{document}